# Fully Automatic Intervertebral Disc Segmentation Using Multimodal 3D U-Net

Chuanbo Wang
Department of Computer Science
University of Wisconsin-Milwaukee
Milwaukee, USA
chuanbo@uwm.edu

Ye Guo
Department of Computer Science
University of Wisconsin-Milwaukee
Milwaukee, USA
yeguo@uwm.edu

Wei Chen
Department of Radiology
Southwestern Hospital
Army Medical University

Zeyun Yu*
Department of Computer Science
University of Wisconsin-Milwaukee
Milwaukee, USA
yuz@uwm.edu

*Abstract*—Intervertebral discs (IVDs), as small joints lying between adjacent vertebrae, have played an important role in pressure buffering and tissue protection. The fully-automatic localization and segmentation of IVDs have been discussed in the literature for many years since they are crucial to spine disease diagnosis and provide quantitative parameters in the treatment. Traditionally hand-crafted features are derived based on image intensities and shape priors to localize and segment IVDs. With the advance of deep learning, various neural network models have gained great success in image analysis including the recognition of intervertebral discs. Particularly, U-Net stands out among other approaches due to its outstanding performance on biomedical images with a relatively small set of training data. This paper proposes a novel convolutional framework based on 3D U-Net to segment IVDs from multi-modality MRI images. We first localize the centers of intervertebral discs in each spine sample and then train the network based on the cropped small volumes centered at the localized intervertebral discs. A detailed comprehensive analysis of the results using various combinations of multi-modalities is presented. Furthermore, experiments conducted on 2D and 3D U-Nets with augmented and non-augmented datasets are demonstrated and compared in terms of Dice coefficient and Hausdorff distance. Our method has proved to be effective with a mean segmentation Dice coefficient of 89.0% and a standard deviation of 1.4%.

## I. Introduction

Intervertebral discs (IVDs) are spine components that provide cushioning between adjacent vertebrae and absorb pressure on the spine. IVD disease is a common condition that affects about 5% of the population in developed countries each year [19], causing pain in the lower back and frequently in the neck and limbs as well. In clinical practice, imaging techniques such as computed tomography (CT) and magnetic resonance imaging (MRI) are widely used for diagnosis and treatment planning of IVD disease. Traditionally, experts manually segment IVDs from medical images. However, this process is tedious and suffers from a disagreement between observers in MRI images [20]. Consequently, research interests in automatic segmentation of IVDs from medical images were captured and such studies can be roughly categorized into two groups: traditional methods and deep learning methods.

Studies in the first group focus on combining computer vision techniques and data-driven machine learning approaches. [22] proposes an IVD segmentation method as an energy minimization problem solved by graph-cuts algorithms where the graph edges are divided into two types: terminal edges that connect the voxels and non-terminal edges that connect neighbor voxels. [23] and [11] apply an atlas-based method that first proposes atlas candidates as initialization and then utilize label fusion to combine IVD atlases to generate the segmentation mask. However, to generate the initialization, [23] registers IVD atlases to the localization obtained by integral channel features and a graphical parts model. Whereas [11] uses data-driven regression to create a probability map, which further defines a region of interest (ROI) as the initialization for segmentation. Another regression-based method [12] is proposed to address the segmentation of multiple anatomic structures in multiple anatomical planes from multiple imaging modalities with a sparse kernel machines-based regression. [24] proposes an automatic method using a conditional random field (CRF) based on super-voxels generated from a variant of simple linear iterative clustering (SLIC). A support vector machine (SVM) is then used to perform super-voxels classification, which is later integrated into the potential function of the CRF for final segmentation using graph cuts. [21] builds an automatic IVD segmentation framework that localizes the vertebral bodies using regression-forests-based landmark localization and optimizes the landmarks by a high-level Markov Random Field (MRF) model of global configurations. The IVD segmentation mask is then generated from an image processing pipeline that optimizes the convex geodesic active contour based on the geometrical similarity to IVDs. In [14], IVD segmentation is performed by iteratively deforming the corresponding average disc model towards the edge of each IVD, in which edge voxels are defined by a 26-dimension feature vector including intensity, gradient orientation and magnitude, self-similarity context (SSC) descriptor, and Canny edge descriptor, etc. These traditional methods have at least one of the following limitations: 1) high computation complexity caused by traditional computer vision techniques in the segmentation pipeline, 2) dependence of segmentation performance on empirically designed parameters and handcrafted features, which do not guarantee optimal result and are not immune to severe pathologies and rare cases, 3) accumulated error propagation through the segmentation pipeline.

Recently, the successes of applying deep learning in the domain of computer vision spark interest in spinal image segmentation using deep convolutional neural networks (CNN) [25]. Traditional machine learning methods typically make decisions based on feature extraction, meaning that, to find the segmentation mask, one must guess which features of the IVDs are important and design sophisticated algorithms that capture these features. However, in CNN, feature extraction and decision making are integrated together. The features are extracted by convolutional kernels and their importance is decided by the network during the training process. A typical CNN architecture consists of convolutional layers and a fully connected layer as the output layer, which requires fixed-size inputs. One successful variant of CNN is fully convolutional neural networks (FCN) [26]. Networks of this type are composed of convolutional layers without any fully connected layer at the end of the network. This enables the network to take arbitrary input sizes and prevent the loss of spatial information caused by the fully connected layers in CNNs. Several FCN-based methods are proposed to solve the IVD segmentation problem. For example, [27] extends the 2D FCN into a 3D version with end-to-end learning and inference. [16] proposes a 3D multi-scale FCN that expands the typical single-path FCN to three pathways where each pathway takes volumetric regions on a different scale. Features from three pathways are then concatenated to generate a probability map, from which the final 3D segmentation mask is generated by simple thresholding.

More recently, U-Net and its variants have outperformed the state-of-art in many biomedical image segmentation tasks, including brain tumor detection [5], liver image segmentation [6], pixel-wise regression [7], etc. The pertinacious architecture and affluent data augmentation allow U-Net to quickly converge to the optimal model from a limited number of annotated samples. Comparing to CNN and FCN, U-Net uses skip connections between contraction and expansion and a concatenation operator instead of a sum, which could provide more local information to global information while expansion. Moreover, U-Net is symmetric such that a large number of feature maps in an expansive path facilitate to transfer more information. U-Net has been widely applied to the IVD segmentation problem. [13] applies a conventional 3D U-Net [15] on the IVD dataset provided by the 3rd MICCAI Challenge [1] of Intervertebral Discs Localization and Segmentation. [28] designs a new network architecture based on U-Net: boundary specific U-Net (BSU). The architecture consists of repeated application of BSU pooling layers and residual blocks, following the idea of residual neural networks (RNN). [18] extends the conventional U-Net by adding three identical pathways in the contracting path to process the multi-modality channels of the input. These pathways are interconnected with hyper-dense connections to better model relationships between different modalities in the multi-modal input images. [17] proposes an IVD segmentation pipeline that first segments the vertebral bodies using a conventional 2D U-Net to find the spine curve and IVD centers. Transverse 2D images and sagittal 3D patches are cropped around the centers to train an RNN fusing both 2D and 3D convolutions. However, the effectiveness of data augmentation and multi-modality input images are not fully explored in these works. Furthermore, the performance of 2D neural networks versus 3D neural networks has remained as an open discussion. Some researchers [18] believe the accuracy of 2D deep learning exceeds that of 3D mainly because the manual segmentations provided in the IVD dataset [2] is usually performed in a slice-by-slice manner in 2D space. These annotation labels contain sharp contours, which makes the segmentation task difficult for 3D CNNs that generally predict smooth contours. However, only considering the pixels in each image from the stack will lose the spatial information between different slices. Since the IVD dataset contains clinical data of lumbar vertebrae and IVDs in the form of 3D volumetric MRI images, other researches [1, 13, 16] apply the 3D convolutional operation.

Our contributions can be summarized as follows: 1) we propose a fully automatic IVD segmentation method using multimodal 3D U-Net by adopting a two-stage pipeline: localizing the IVDs followed by segmenting IVDs based on the localization; 2) we propose to deal with multi-modality data in a new perspective by analyzing the differences between various modalities with respect to image properties of the input data and examining the effectiveness of different combinations of modalities based on our analysis in the conducted experiments; 3) we run experiments to explore the superiority of 3D U-Net over 2D U-Net even though a consistent agreement has not been reached yet [1, 13, 18]. The rest of the paper is structured as follows: Section II introduces methods we use to tackle the above-mentioned problems in detail. Section III presents the quantitative and visual evaluation of the methods. Section IV concludes our work.

II. METHODS

In this work, we use two U-Net-based models (2D and 3D) to demonstrate their difference in effectiveness to tackle the IVD segmentation problem. Since the clinical data are multi-modality images, we first analyze these data and decide how to effectively integrate partial or all modalities to generate more accurate segmentation results. Then augmentation techniques, including homogeneous and non-homogeneous deformations, are utilized to increase the number of training samples. For the 3D framework, we apply 3D convolutional kernels directly on the volume and generate a volumetric mask using a modified multimodal 3D U-Net. For the 2D framework, we train a modified 2D U-Net that takes images slices from the volumetric dataset as input and predict the IVD segmentation mask in 2D space. In the following sections, we introduce the dataset, the multi-modality analysis of the dataset and our augmentation method. Then the architectures of our 2D network and 3D network are presented in detail.

## A. Dataset

The dataset [2], by courtesy of Prof. Guoyan Zheng from University of Bern, consists of 8 sets of 3D multi-modality MRI spine images collected from 8 patients in 2 different stages of prolonged bed test. Each spine image contains at least 7 IVDs of the lower spine (T1-L5) and four modalities following Dixon protocol: in-phase (inn), opposed-phase (opp), fat and water (wat) images. In detail, water images are spin echo images acquired from water signals. fat images are spin echo images acquired from water signals. In-phase images are the sum of water images and fat images. Opposed-phase images are the difference between water images and fat images. In total, there are 32 3D single-modality volumes and 66 IVDs. For each IVD, segmentation ground truth is composed of binary masks manually labeled by three trained raters under the guidance of clinicians.

## B. Multi-modality Analysis

The traditional multi-modality deep learning methods conventionally fed all modalities images into different channels as input to the neural network. Zhang et. al. use multi-modality images to segment infant brains [16]. Li et. al. trained the 3D FCN separately on every single modality (fat, in-phase, opposed-phase, and water) and then on a merged full-modality of the spine dataset to validate the superiority of training with multi-modality data. Both works show that training on the full-modality images could yield more accurate IVD segmentation results than the single-modality strategy [17]. Moreover, Li's research indicates that the single-modality of opposed-phase images and water images could enhance the performance than using the modality of fat images and in-phase images. Nonetheless, the full-modality-fused images cannot guarantee to achieve its full capacity efficiently in all experiments. The dependency between diverse modalities could lead to data co-adaption, which means the same features are detected repeatedly. In addition, the significant difference between modalities at the same locations could trigger data corruption problem to misdirect the corresponding neurons.

To alleviate these problems, we made a further exploration on the intensity distribution of various modalities and analyzed the mean gradient on the boundary of the intervertebral discs. As shown in Fig. 1, the images of fat modality have obviously low contrast at the edges compared with the images of the other three modalities. In more detail, assume the target intervertebral discs are foreground and the other areas are background and Table I lists the average intensities and the standard deviations for the foreground and background samples using the provided label maps as a reference. Moreover, Table I also shows the absolute Weber contrast coefficients of in-phase, opposed-phase, fat and water images [3], which is computed by:

$$C = \frac{|I - I_b|}{I_b}$$

TABLE I. Absolute Weber contrast

| Modality | Mean ± SD Foreground Intensity | Mean ± SD Background Intensity | Absolute Weber Contrast |
|---|---|---|---|
| fat | 15.9 ± 12.7 | 35.2 ± 47.5 | 0.57 |
| in-phase | 172.1 ± 39.9 | 97.9 ± 89.5 | 0.75 |
| opposed-phase | 155.4 ± 47.5 | 63.2 ± 65.1 | 1.46 |
| water | 163.4 ± 41.1 | 67.9 ± 67.5 | 1.43 |

where $I$ is the mean intensity value of the foreground voxels and $I_b$ is the mean intensity value of the voxels in the background. Comparing to other modalities, fat and in-phase images have relatively low Weber contrast values. To take full advantage of multiple modalities of this dataset, we take a step further from [16] and train our network on different combinations of multi-modality images to examine the effectiveness of the fat and in-phase images. The comprehensive results will be presented later in the results section.

## C. Data Augmentation

When only a few training samples are available, data augmentation is essential to increase the size of the desired dataset and increase the network's robustness to data variances. With only 8 sets of spine data provided, our networks demand more data for training. Especially for the 2D U-net, augmenting the training samples could significantly enhance the performance. A variety of conventional 3D image processing techniques such as translations, rotations, flip, and scaling are adopted in our method. In addition to these affine transformations, as a particularly important augmentation technique in biomedical segmentation tasks, elastic deformation is also used in combination with other transform functions due to the fact that the most common variation in tissue is deformation [2]. We fabricate smoothly deformed surfaces by generating random displacement fields and the fields are convolved with a Gaussian of standard deviation δ (in pixels). And the displacement fields are then multiplied by a scaling factor α that controls the intensity of the deformation [2]. The per-pixel displacements are computed using bicubic interpolation. Based on the augmentation functions mentioned above, a random combination of operations is selected in an arbitrary order and applied to the original spine dataset. Consequently, the size of the training dataset is increased from 6 to 24 (2 of the 8 sets of original spine data are not used for augmentation but for validation of the prediction). To sufficiently take advantage of the information between adjacent layers, besides using the augmented spine data in 3D framework, we also use the stack of images sliced along each axis of the 3D augmented data as the training set of the 2D network rather than directly applying 2D augmentation techniques on the sliced images from the original dataset.

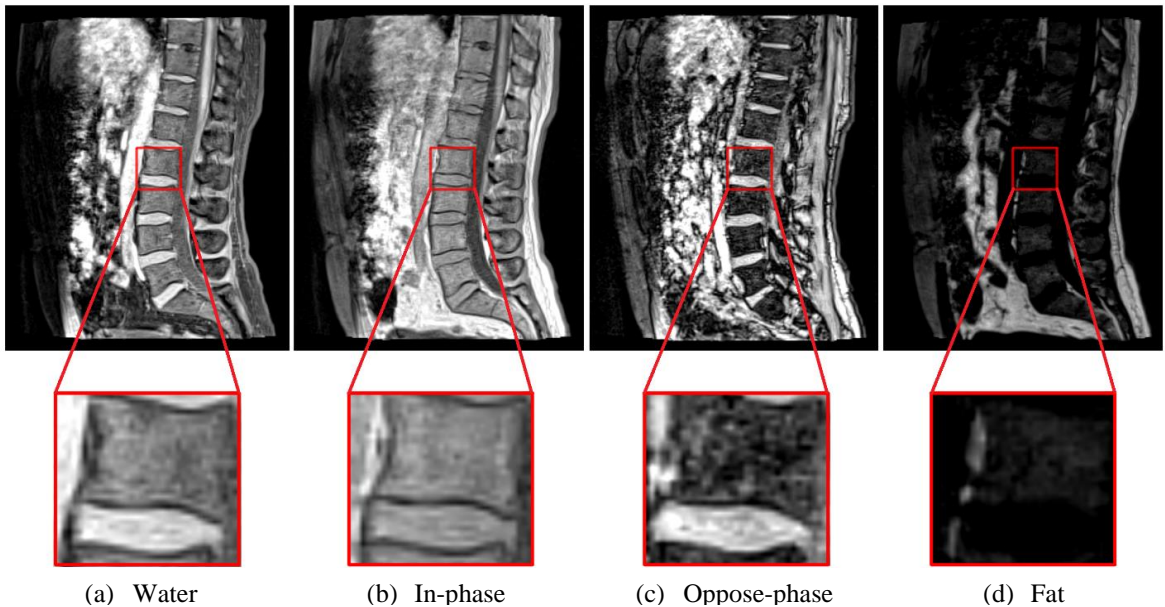

| (a) Water | (b) In-phase | (c) Oppose-phase | (d) Fat |

Fig. 1. The comparison of contrast in different modality of images (The modalities from left to right are respectively fat, opposed-phase, in-phase, water).

## D. Training

Both 2D and 3D networks in the presented work are implemented in python with Keras [9] and Tensorflow [10] backend. Batch size is set to 1 for the 3D network and 16 for the 2D network to obtain optimized training accuracy. The models are trained on a PC with an 8-core 3.4GHz CPU and a single NVIDIA GTX TITAN XP GPU. The training time of a single epoch takes about 3 s in the 3D model and 45 s in the 2D model. Eventually, the validation loss will stop increasing at around 10000 epochs for the 3D model and 500 epochs for the 2D model before overfitting.

The convolutional kernels of our networks are initialized with the HE initialization [29] to speed up the training process. For updating the parameters in the networks, we employ the Adam optimization algorithm [30], which has been popularized in the field of stochastic optimization due to its fast convergence compared to other optimization functions. The learning rate is set to 1e-5 for accurate predictions and reasonable training time. The energy function is computed by the Dice-coefficient loss function defined as:

$$E = \frac{2 \times I + S}{\sum T + \sum P + S}$$

where T stands for the ground truth label and P for the prediction. $I$ is the number of true positives calculated by the number of intersection pixels between $T$ and $P$. $S$ is a smoothing factor with a value of 1.

## E. 2D Network

The soft tube models of the lumbar spine column are all 3D volumetric data such that partitioning the same 3D model from different angles could create different 2D image datasets. We train the 2D U-Net on image slices extracted from the augmented spine dataset along varying directions (x-axis, y-axis, and z-axis) to predict segmentation masks. The stack of images sliced along the x-axis is of size 256x256 pixels while the images retrieved from the other two axes are of size 256x36 pixels. To make the trained network handle images of different sizes, we preprocess all images to the same size of 256x256 pixels using zero-padding. It's noted that the number of images retrieved from a single 3D sample along y-axis or z-axis is 256 while the number of images sliced from x-axis direction decreases to 36, which is roughly seventh of 256. As a result, the training set based on y-axis or z-axis is six times bigger than that on the x-axis. More experimental results and comparisons are provided in Section III.

The architecture of our modified 2D U-Net is shown in Fig. 2. The contracting path consists of 4 convolutional blocks. Each block contains two convolutional layers with a filter size of 3×3, stride of 1 in both directions, same zero paddings and rectifier activation, which increase the number of feature maps from 1 to 1024. For the down-sampling, max pooling with stride 2×2 is applied to the end of every block except the last block, so the size of feature maps decreases from 512×512 to 32×32. In the expansive path, every block starts with a deconvolutional layer with a filter size of 3×3 and stride of 2×2, which doubles the size of feature maps in both directions but decreases the number of feature maps by two, so the size of feature maps increases from 32×32 to 512×512. In every expansive block, two convolutional layers reduce the number of feature maps of the concatenation of deconvolutional feature maps and the feature maps from encoding path.

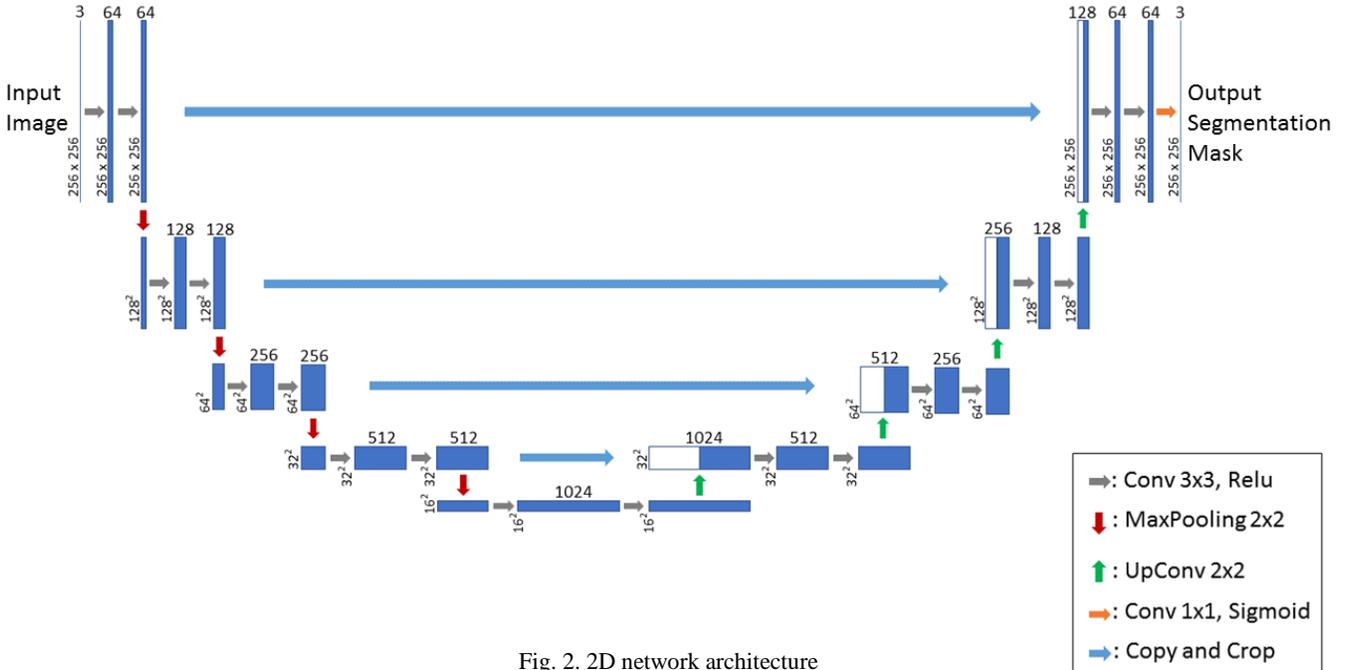

Fig. 2. 2D network architecture

*F. 3D Network*

In our proposed 3D method, a two-stage coarse-to-fine strategy is used to tackle the segmentation problem directly on 3D volumes. The general workflow is illustrated in Fig. 3. In the first stage, each IVD is localized and a voxel is assigned as its center. These centers are used to divide the volume into small 3D patches, each of which contains a single IVD. In the second stage, a multimodal deep learning model is trained on the patches for precise segmentation.

Medical images are often complex and noisy in nature where the region of interest (ROI) is relatively small comparing to the background. We first localize the IVDs in the image and then crop 3D patches based on the localization. This not only gets rid of some background but simplifies the problem for the segmentation stage and reduces the computational cost as well. It has been shown that 3D U-Net achieves the best localization result but not the best segmentation result [1]. We use this two-stage strategy to work around this problem. In the end, post-processing is performed to generate the final segmentation.

*1) Localization*

For the localization of IVDs, we train a localization network, which is a conventional 3D U-net, on the IVD dataset to roughly locate the IVDs from the volume. From this segmentation, we have a good estimate of IVD centers by finding the center of each connected component after removing small regions.

From our observation, IVDs are generally sparsely located in 3D space with a distance from each other and share a common disc-like morphological profile. Thus, we simply put a 35*35*25 bounding box around each estimated center to crop a 3D patch. Then we zero-pad the patches to 36*36*28 so they can be nicely fed into the segmentation network in the next stage described below.

*2) Segmentation*

For IVD segmentation from the 3D patches, we employ a modified 3D U-Net architecture shown in table II that essentially looks at IVD segmentation as a regression problem. This network takes 3D patches as input and predicts 3D patches where the intensity value on each voxel stands for how confident is the network in the voxel belonging to an IVD. Table II presents an overview of the architecture of our 3D segmentation network. Each step in the contracting path consists of repeated application of two 3x3x3 unpadded 3D convolutions followed by a rectified linear unit (ReLU). A dropout operation is inserted between the two convolutions to reduce the dependence on the training dataset and increase the accuracy. A dropout rate of 0.2 is used following the analysis [8] on the dropout effect in CNN. We also apply batch normalization to speed up and stabilize the training process and a 2x2x2 max pooling layer with stride 2 for down-sampling after every two convolutional layers. At each down-sampling step, we double the number of feature channels. Every step in the expansive path consists of an up-sampling of the feature map followed by a 2x2x2 up convolution that halves the number of feature channels, a concatenation with the corresponding feature map from the contracting path, and two 3x3x3 convolutions, each followed by a ReLU. The output layer is a 1x1x1 convolution layer with sigmoid activation used to generate the segmentation mask for each modality. In total the network has 12 convolutional layers and 1.4 million parameters.

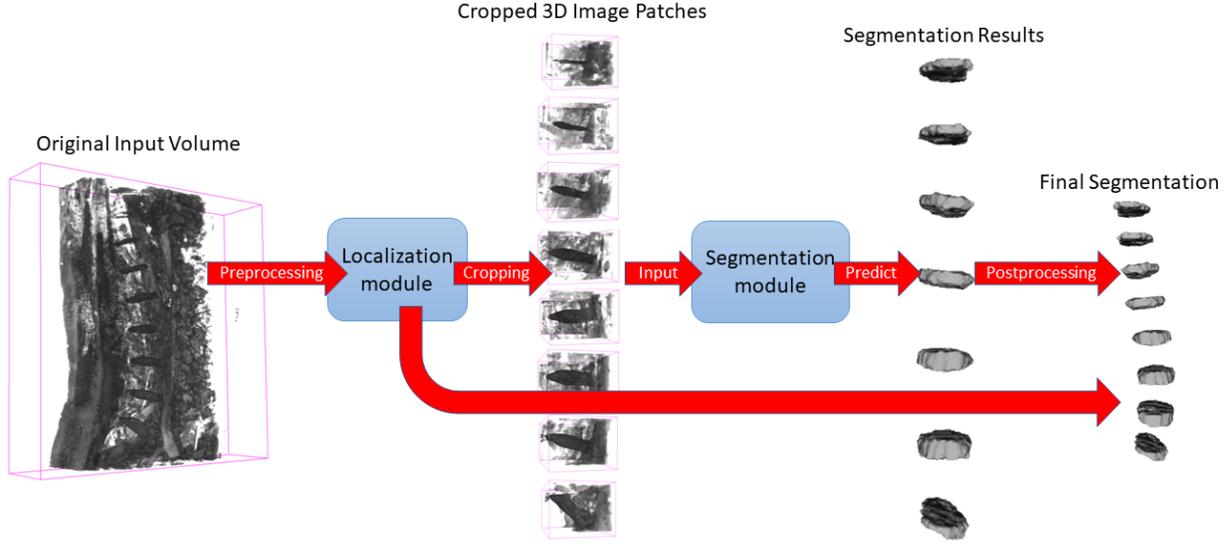

Fig. 3. Workflow of the proposed 3D method

TABLE II. 3D Segmentation Network Architecture

|  | Layer name | Input shape | Output shape |
|---|---|---|---|
|  | Input | 36×36×28×3 | 36×36×28×3 |
|  | Conv3D | 36×36×28×3 | 36×36×28×32 |
|  | Dropout | 36×36×28×32 | 36×36×28×32 |
|  | Conv3D | 36×36×28×32 | 36×36×28×32 |
|  | Max Pooling | 36×36×28×32 | 18×18×14×64 |
| Contracting Path | Conv3D | 18×18×14×64 | 18×18×14×64 |
|  | Dropout | 18×18×14×64 | 18×18×14×64 |
|  | Conv3D | 18×18×14×64 | 18×18×14×64 |
|  | Max Pooling | 18×18×14×64 | 9×9×7×128 |
|  | Conv3D | 9×9×7×128 | 9×9×7×128 |
|  | Dropout | 9×9×7×128 | 9×9×7×128 |
|  | Conv3D | 9×9×7×128 | 9×9×7×128 |
|  | Up Sampling | 9×9×7×128 | 18×18×14×64 |
|  | Concatenate | 18×18×14×64 | 18×18×14×64 |
|  | Conv3D | 18×18×14×64 | 18×18×14×64 |
|  | Dropout | 18×18×14×64 | 18×18×14×64 |
|  | Conv3D | 18×18×14×64 | 18×18×14×64 |
| Expensive Path | Up Sampling | 18×18×14×64 | 36×36×28×32 |
|  | Concatenate | 36×36×28×32 | 36×36×28×32 |
|  | Conv3D | 36×36×28×32 | 36×36×28×32 |
|  | Dropout | 36×36×28×32 | 36×36×28×32 |
|  | Conv3D | 36×36×28×32 | 36×36×28×32 |
|  | Conv3D | 36×36×28×32 | 36×36×28×2 |
|  | Batch Norm | 36×36×28×2 | 36×36×28×2 |
|  | Output | 36×36×28×2 | 36×36×28×3 |

*3) Post-processing*

The prediction from the segmentation stage contains 3D patches with continuous voxel intensity values that representing how confident is the network in the voxel belonging to an IVD. The final segmentation mask for each patch is obtained by binary thresholding with a threshold of 0.5, which means voxels that are predicted more likely to be IVD voxels than background voxels are included in the segmentation mask. Then the mask patches are assembled back to a 3D volume of the lower spine, with the same size of the IVD dataset, using the IVD center locations from the localization stage and zero-padding.

## III. RESULTS

In this chapter, we compare and analyze the segmentation performance of our 3D method with the state-of-art methods. Our 3D deep learning model is trained with data augmentation and preprocessing. To validate data augmentation, we train our baseline deep learning model on the original dataset without any data augmentation and compare its segmentation performance with our 3D method. Extensive experiments are conducted to investigate the effectiveness of multimodal MRI image data by training our 3D segmentation network on the dataset consisting of different combinations of image modalities. Our 2D network is also compared with the 3D method to explore which one is superior in the field of 3D spinal image segmentation. For fairness of comparison, we use the same training strategies and data augmentation strategies throughout the experiments.

*A. Evaluation metrics*

To evaluate the segmentation performance, two metrics are adopted from the 2015 MICCAI Challenge [1]:

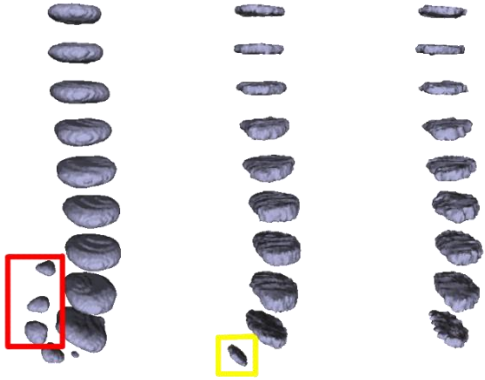

Fig. 4. Training without in-phase images.
Left: segmentation results of the baseline model trained with all modalities in the dataset. In the middle: segmentation results trained without in-phase images. Right: the ground true label.

TABLE III: Segmentation performance of our 3D method using different combinations of modalities as the training dataset

| Training dataset | Combination | Mean Dice ± SD |
|---|---|---|
| 1) | opp, wat, fat, and inn | 87.9 ± 1.7 |
| 2) | opp, wat, and fat | 89.0 ± 1.4 |
| 3) | opp, wat, and inn | 88.0 ± 1.6 |
| 4) | opp, and wat | 88.5 ± 1.6 |

*Dice coefficient (Dice)*: Dice shows the similarity between the segmentation and the ground truth. More specifically, Dice measures the rate of correctly segmented voxels that is computed by:

$$Dice = \frac{2 \times |X \bigcap Y|}{|X| + |Y|} \times 100\%$$

where $X$ denotes the set of voxels in the ground truth segmentation mask and $Y$ is the set of voxels segmented by our method. A larger Dice score indicates a more accurate segmentation result.

*Hausdorff distance (HD)*: HD measures the Hausdorff distance between two surface meshes. We compute HD for surfaces reconstructed from the ground true segmentation mask and our segmentation result. Surfaces are generated using Iso2mesh [4] from binary segmentation masks. The closest distance from each vertex on the source surface mesh to the target surface mesh is found and HD is then computed. A smaller HD value indicates better segmentation performance.

### B. Effectiveness of the multi-modality data

As shown in Fig. 4, the segmentation results achieved by excluding the in-phase images from the training dataset are more accurate and less noisy near the lower IVDs than that by the original full-modality data. Moreover, using the training dataset without in-phase images, our localization network is able to learn much more details and make much more accurate predictions about the IVD centers. This makes the localization of centers more stable and allows us to simply remove small regions (marked by yellow boxes) and then crop a fix-size 3D patch for each IVD in the volume to train the segmentation network.

From the multi-modality analysis, we found that the fat and in-phase images have a significantly lower contrast among all the modalities. To analyze the effectiveness of the multi-modality input data, we train our 3D network on 4 different combinations of input modalities: 1) we train the network on full-modality images as the baseline, 2) we exclude the fat images from all 4 modalities to build the second training dataset, 3) the fat images are excluded from all 4 modalities, and 4) we only include oppose-phase and water images in the last training dataset. The mean Dice scores of the segmentation results predicted by the network trained on each dataset are presented in Table III. Among all the different training settings, the network trained on full-modality images shows the worst segmentation performance. The reason is that the fat and in-phase images have a lower contrast, which means that the input values of the network are closer to each other and make it more difficult for the convolutional kernels to distinguish between them. It is worthy of pointing out that input normalization does not help with this situation because it is performed over the values of all the modalities. In other words, if we treat these 4 types of images equally during the training process, the fat and in-phase images confuse the network with their low image contrast.

### C. Comparison of our methods and state-of-the-art methods

To evaluate the performance of the proposed methods, we compare the segmentation results achieved by our methods with those by 3D U-Net[15], the CNN-based team UNICHK [13] and the winning team UNIJLU [14] in the test1 dataset of the 2015 MICCAI Challenge [1]. To investigate the effectiveness of 3D implementations over 2D, we compare our 2D network and 3D network with respect to segmentation performance. We also conduct experiments to explore the effectiveness of training our 2D network using the images sliced along different directions of the volumetric data. A demonstration of segmentation results from our 2D and 3D method can be found in Fig. 5 where we select 4 continuous slices (from the $13^{th}$ slice to the $16^{th}$ slice) that present all 9 IVDs in the sagittal view.

Quantitative results evaluated with the different architectures are presented in Table IV. The mean Dice score obtained by our 3D method is 89.0% with a standard deviation (SD) of 1.4%. We bring a 1.5% boost comparing to the conventional 3D U-Net by training our network on 3D image patches extracted from opposed-phase, water, and fat images. This result is still 2.5% behind the state-of-the-art performance achieved by UNIJLU. The Mean HD of our 3D method reached 0.8 mm with an SD of 0.3 mm, which indicates that our method is slightly better when the segmentation results are reconstructed to 3D models. The strength of deep learning methods is the computation time. The Theano-based implementation of 3D U-Net from UNICHK takes 3.1 s to process one 40 × 512 × 512 volume. Our network is implemented based on Tensorflow and it takes about 0.5 s to segment all the IVDs in a 36 × 256 × 256 input volume. Overall, the computation time of our end-to-end segmentation is about 10 s including localization, preprocessing, segmentation and postprocessing. Whereas it takes 5 min on average to segment all IVDs for a patient by UNIJLU. It is also worth mentioning that the training dataset used in our study only contains data from 6 patients while UNICHK and UNIJLU have access to a training dataset from 16 patients i.e. our network is able to learn the 3D geometric morphometrics of IVDs with much less data to learn from.

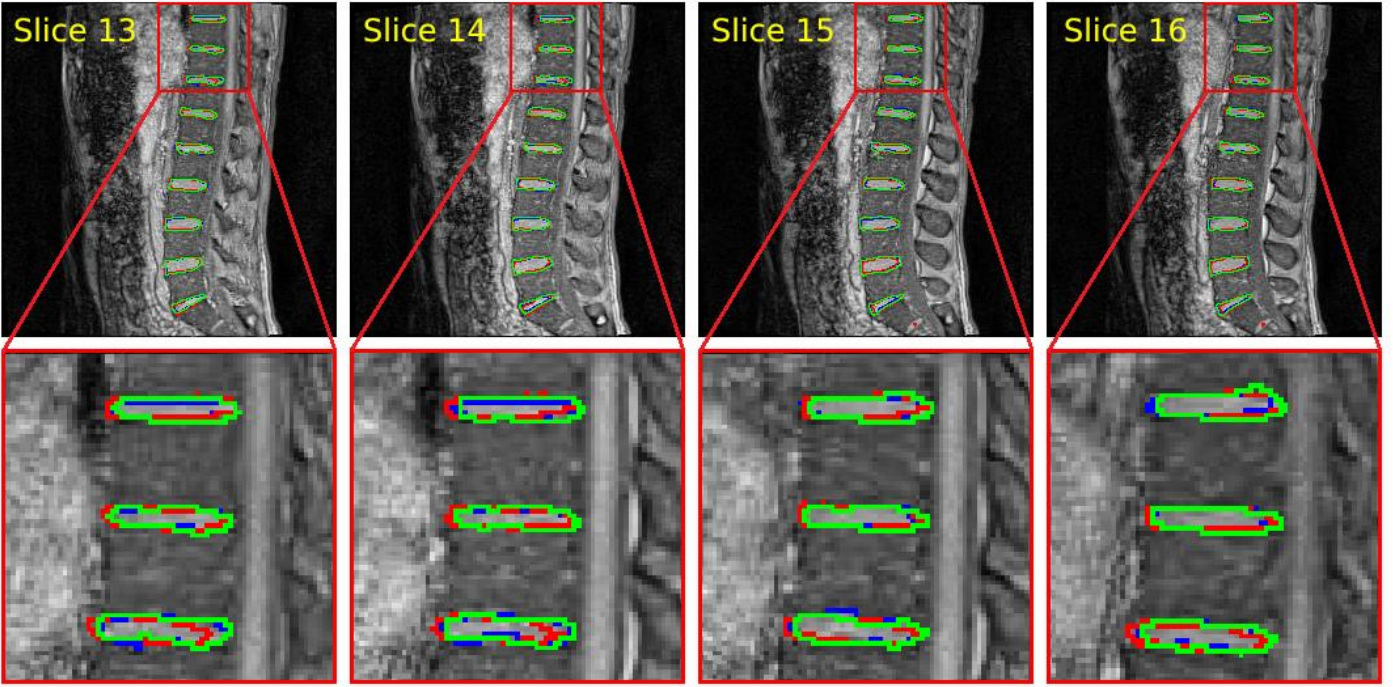

Fig. 5: A comparison between our 2D and 2D method. Segmentation masks of our 2D method are mapped into 3D space and marked by red contours. Segmentation masks of our 3D method are marked by green contours. Ground true labels are marked by blue contours.

We observe that our 3D implementation achieves much better segmentation result compared to the 2D version. The reason is that 2D convolutional filters do not take the spatial relationship between slices into consideration. Even though raters labeled the training dataset in a slice-by-slice manner, they might make decisions accordingly with the adjacent slices in mind. This result agrees with the conclusion stated in [1, 13].

Using the sliced images from different directions of the volumetric data will produce different accuracies in the training of 2D U-Net. Table V lists the corresponding prediction performances based on the datasets obtained by partitioning the augmented spine columns from the x-axis, y-axis, and z-axis, respectively. Moreover, we overlap the models generated from the three axes and take the average intensity as the value for each voxel and then use half of the maximum value, which is 127.5, as the threshold to distinguish truth and background. All experiments use two sets of original labeled spine data as input. The results reveal that the segmented IVDs from y-axis have a relatively higher mean accuracy than from x-axis and z-axis and the trained model is more sensitive to the inputs of different qualities. Furthermore, using the average intensity from all directions could not improve the outcomes.

TABLE IV. Segmentation result evaluation of the conventional 3D U-Net, UNICHK, UNIJLU and our methods.

| Methods | Mean Dice ± SD | Mean HD ± SD |
|---|---|---|
| 3D U-Net | 87.5 ± 0.9 | 1.1 ± 0.2 |
| UNICHK | 88.4 ± 3.7 | 1.3 ± 0.2 |
| UNIJLU | 91.5 ± 2.3 | 1.1 ± 0.2 |
| Our 3D method | 89.0 ± 1.4 | 0.8 ± 0.3 |
| Our 2D method | 81.8 ± 1.3 | 2.4 ± 1.0 |

TABLE V. Segmentation performance of the 2D U-Net trained with images sliced in X, Y, and Z direction, respectively.

| Directions | Mean Dice ± SD |
|---|---|
| X-Axis | 79.4 ± 1.0 |
| Y-Axis | 81.8 ± 1.3 |
| Z-Axis | 77.3 ± 67 |
| Average of All Axes | 80.0 ± 1.5 |

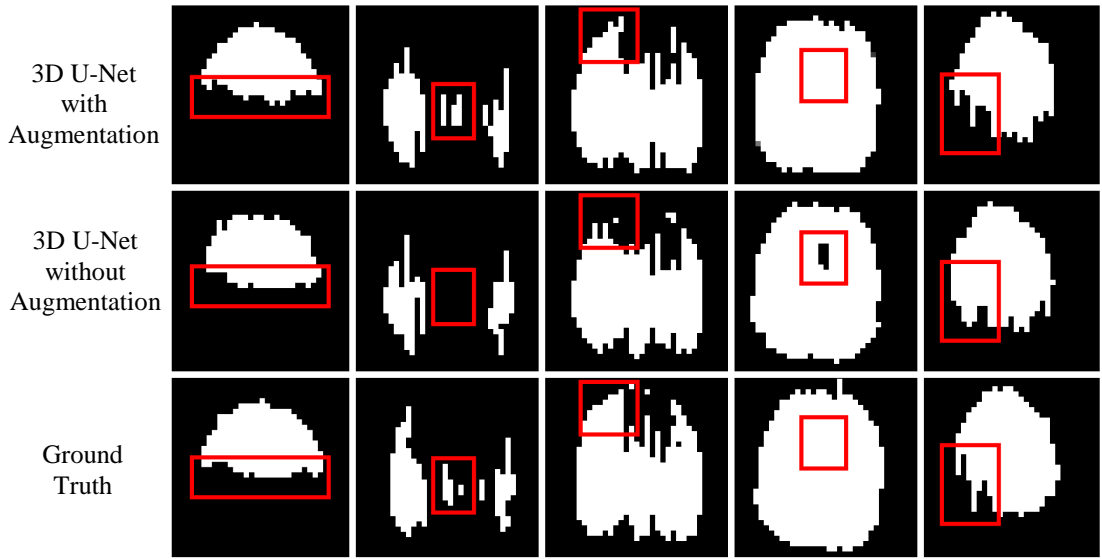

Fig. 6: Effectiveness of data augmentation. Each column represents a MRI slice. The segmentation mask predicted by 3D U-Net with data augmentation, 3D U-Net without data augmentation and the ground truth label of the slice are shown, respectively, in the transverse view.

*D. Effectiveness of Data Augmentation*

To explore the effectiveness of data augmentation when applying 3D CNNs to IVD segmentation, we conducted experiments by training our 3D network with and without the proposed data augmentation method. The results do not suggest obvious differences in terms of Dice and HD. However, we found that data augmentation enables the network to learn more details in the boundary area. More specifically, segmentation results are improved in the regions between an IVD and the adjacent vertebral body. Examples are presented in Fig. 6, the regions well segmented by our 3D network trained with augmented dataset are marked by red boxes. Segmentation on these regions with sharp boundaries is a difficult task for convolutional kernels since CNNs tend to predict smooth contour. The core of an IVD is composed of jelly-like material and the smooth elastic deformation technique in our data augmentation mimics the real-world deformation of IVDs and enriches the training dataset by adding more morphological variants. This makes CNNs more capable to deal with unseen IVD data and make better predictions.

## IV. CONCLUSIONS

In this paper, we attempt to solve the automated segmentation problem of intervertebral discs from two different perspectives, one based on 2D image data and the other based on 3D volumetric data. Among all various types of deep neural networks, U-Net has demonstrated its superiority in the field of biomedical image identification due to the massive data augmentation and the special synonymous fully convolutional architecture. The up-sampling part which consists of a large number of feature channels allows the network to propagate localization combined with contextual information to higher resolution layers. We first analyze the influences of single modality and multi-modalities to the prediction outcomes and find out that using the combination of the modalities of fat, opposed-phase, and water will yield more satisfying results than using other combinations. Then the 2D U-Net is implemented to create the predictions map based on the image slices retrieved from various directions. Furthermore, a novel approach based on 3D U-Net using cropped 3D image patches as the training dataset is proposed. These patches are identical cuboids centered at the center of each localized intervertebral disc. This approach can achieve a mean accuracy of 89.0% which is above that of most well-known methods in the field.

In the future, we plan to improve our work by integrating the spine-curve-based augmentation and using adaptive window size to produce the 3D image patches. Also, we will try to include more patient data in the training set to improve the robustness and prediction accuracy of our network.